\documentclass[12pt]{article}
\font\sqi=cmssq8
\def\DR{\rm I\kern-1.45pt\rm R}
\def\DC{\kern2pt {\hbox{\sqi I}}\kern-4.2pt\rm C}
\newcommand{\ben}{\begin{enumerate}}
\newcommand{\een}{\end{enumerate}}
\newcommand{\beq}{\begin{equation}}
\newcommand{\eeq}{\end{equation}}
\newcommand{\bse}{\begin{subequation}}
\newcommand{\ese}{\end{subequation}}
\newcommand{\bea}{\begin{eqnarray}}
\newcommand{\eea}{\end{eqnarray}}
\newcommand{\bc}{\begin{center}}
\newcommand{\ec}{\end{center}}

\def\r{r_0}

\begin{document}\begin{center}
{\bf \Large Supersymmetric K\"ahler oscillator  in a constant magnetic field.}\\
\vspace{0.5 cm}
{\large Stefano Bellucci and Armen Nersessian}
\end{center}
$\;^1$ {\it INFN, Laboratori Nazionali di  Frascati, P.O. Box 13,
I-00044 Frascati, Italy}\\
$\;^2$ {\it Yerevan State University, A.Manoogian, 1, Yerevan,
375025 Armenia}\\
$\;^3$ {\it Yerevan Physics Institute, Alikhanian Brothers St., 2,
Yerevan, 375036,
 Armenia}
\begin{abstract}
We propose   the notion of the oscillator on K\"ahler space and consider its
 supersymmetrization in the presence of a constant  magnetic field.
\end{abstract}
Supersymmetric mechanics attracts permanent interest
since its  introduction \cite{witten}. However,
studies focussed mainly on the  mechanics with standard ${\cal N}=2$ supersymmetries
(see for the review \cite{sukh} and refs therein). The systems with
${\cal N}=4$ supersymmetries also received much attention:
the most general
 ${\cal N}=4, D=1,3$  supersymmetric mechanics described by real superfield
actions  were  studied in Refs. \cite{ikp,is} respectively, and those
in arbitrary $D$ in Ref.\cite{dpt};
in \cite{bp} ${\cal N}=4, D=2$ supersymmetric mechanics described
by chiral superfield actions were considered. Let us mention also some recent papers on this subject \cite{recent}.
The study of ${\cal N}=8$ supersymmetric mechanics has been
performed recently in Ref.\cite{n8}.

On the other hand, not enough attention has been paid to systems with non-standard supersymmetry algebra,
although they arise in many realistic situations. Some of the systems of that sort were extensively studied by
M. Plyushchay \cite{plyushchay}. A. Smilga
studied the dynamical aspects of  ``weak  supersymmetry" \cite{smilga} on the
simple example of the supersymmetric oscillator.
 He suggested in this case a nontrivial model of
``weak supersymmetric" mechanics, related with quasi-exactly solvable models
and the systems with nonlinear supersymmetry.\\

In the present  work we consider the supersymmetrization of a specific model of
Hamiltonian mechanics  on K\"ahler manifold
$(M_0, g_{a\bar b}dz^ad{\bar z}^{\bar b})$ interacting with constant magnetic field $B$, viz
\beq
{\cal H}=g^{a\bar b}(\pi_a{\bar \pi}_b+
\omega^2\partial_a K {\bar\partial}_b K),\quad \Omega_0=
d\pi_a\wedge dz^a+ d{\bar\pi}_a\wedge d{\bar z}^a
+iBg_{a{\bar b}} dz^a\wedge d{\bar z}^b,
\label{0}\eeq
where $K(z,\bar z)$ is a K\"ahler potential of configuration space.

Notice, that  the K\"ahler potential  is defined up to holomorphic and antiholomorphic terms,
\beq
K(z,\bar z)\to K(z, \bar z)+ U(z)+ \bar U (\bar z)\; ,
\eeq
while the Hamiltonian  under consideration is not invariant under these transformations.
For example, in the limit $\omega\to 0$ it yields the Hamiltonian
\beq
{\cal H}=g^{a\bar b}(\pi_a{\bar \pi}_b+
\partial_a U(z) {\bar\partial}_b \bar U(\bar z)).
\eeq
This Hamiltonian  admits, in the absence of magnetic field,
 a ${\cal N}=4$ superextension \cite{n4}, in the spirit of Alvarez-Gaum\'e-Freedman \cite{agf}.

The  suggested system could be viewed, in many cases,   as a generalization of the
oscillator on the K\"ahler manifold.
It includes, as special cases, a few interesting exactly-solvable systems.
\begin{itemize}
\item
The oscillator on $\DC^n= \DR^{2n}$,
\beq
{\cal H}=\pi {\bar \pi}+\omega^2 z\bar z ,
\eeq
corresponding to the choice $U=z^az^a/2$.
The  constants of motion defining the hidden symmetries of the system, could be represented as follows:
\beq
I^+_{ab}=\pi_a\pi_b+\omega^2\bar z^a \bar z^b, \quad I^-={\bar I}^+,\quad
I_{a\bar b}=\pi_a\bar\pi_b+\omega^2\bar z^a  z^b.
\eeq
The symmetry  algebra of the system is $u(2n)$.
Clearly, these constants of motion are functionally-dependent ones.

\item The oscillator on complex
projective space $\DC P^n$ (for $n>1$) \cite{cpn},
\beq
K=\r^2\log(1+z\bar z), \Rightarrow {\cal H}=g^{\bar a b}\bar\pi_a\pi_b+\omega^2\r^2 z\bar z.
\eeq
This system  is also specified, in the absence of magnetic field,  by the hidden symmetry
 given by the constants of motion
\beq
J_{a\bar b}={i}(z^b\pi_a-\bar\pi_b\bar z^a), \quad
I_{a\bar b}=
\frac{J^+_a J^-_b}{\r^2} +\omega^2\r^2 {\bar z}^a z^b\; ,
\label{sym}\eeq
where $J^+_a=\pi_a+(\bar z\bar\pi)\bar z^a$, $J_a^-=\bar J^+_a$ are the translation generators.

These generators form the   nonlinear (quadratic) algebra
\beq
\begin{array}{c}
\{J_{{\bar a} b}, J_{\bar c d}\}=
i\delta_{\bar a d}J_{\bar b c}
-i\delta_{\bar c b}J_{\bar a d},\quad
\{I_{a\bar b}, J_{c\bar d}\}=
i\delta_{c\bar b}I_{a\bar d}-i\delta_{a\bar d}I_{c\bar b} \\
 \{I_{a \bar b}, I_{c\bar d}\}=
i\omega^2 \delta_{c\bar b} J_{a\bar d}- i\omega^2
 \delta_{a\bar d}J_{c\bar b}

+i I_{c\bar b}(J_{a\bar d}+J_0\delta_{a\bar d})/\r^2
- i I_{a\bar d}(J_{c\bar b}+J_0\delta_{c\bar b})/\r^2 \quad .
\end{array}
\label{cpnalg}\eeq

\item
The oscillator on $\DC P^2$ could also be extended to the  class of K\"ahler conifolds,
defined by the K\"ahler potential \cite{eran}
\beq
K=\r^2\log(1 \pm (z\bar z)^{\nu^2}) \; ,\quad  \Rightarrow\quad
{\cal H}=g^{\bar a b}\bar\pi_a\pi_b+\omega^2\r^2( z\bar z)^{\nu^2},
\eeq
where $\nu$ is a numerical parameter.
Although the corresponding oscillator systems do not have hidden symmetry for $\nu\neq 1$,
i.e. on non-constant curvature spaces, they remain exactly-solvable at both  the classical \cite{eran}
and the quantum \cite{ben} level.
Moreover, for $\nu=2$ the conic oscillator reduces to the Higgs oscillator on the three-dimensional
 sphere  and pseudosphere  interacting with a Dirac monopole field and
 some specific potential proportional to the squared monopole number.\\
\end{itemize}
Notice also , that the Hamiltonian system under consideration, yields, in the ``large mass limit",
$\pi_a\to 0$, the following one,
$${\cal H}_0=\omega^2 g^{a\bar b}\partial_a K {\bar\partial}_b K,\quad
\Omega_0= iBg_{a\bar b}dz^a\wedge d\bar z^b,$$
which could be easily extended with ${\cal N}=2$ supersymmetry \cite{npps}.

 The supersymmetrization procedure follows closely the steps we performed
 in \cite{cpn} for the oscillator on complex projective space.
 Next we will show that, although the system under
 consideration does not possess a standard ${\cal N}=4$
 superextension, it admits  a superextension in terms of a nonstandard superalgebra with four fermionic generators,
 including, as subalgebras, two copies of ${\cal N}=2$ superalgebras. This nonstandard superextension
 respects the inclusion of constant magnetic field.

We  follow the following strategy. At first, we extend the initial phase space to the
 a $(2N.2N)_{\DC}$-dimensional  superspace
 equipped with the symplectic structure
\begin{equation}
\begin{array}{c}
\Omega=d\pi_a\wedge dz^a+ d{\bar\pi}_a\wedge d{\bar z}^a +i(B
g_{a\bar b}+iR_{a{\bar b}c\bar d}\eta^c_\alpha\bar\eta^d_\alpha)
dz^a\wedge d{\bar z}^b+ g_{a\bar
b}D\eta^a_\alpha\wedge{D{\bar\eta}^b_\alpha}\quad .
\end{array}
\label{ss}\end{equation} Here $D\eta^a_\alpha
=d\eta^a_\alpha+\Gamma^a_{bc}\eta^a_\alpha dz^a, \quad
\alpha=1,2$, and
 $\Gamma^a_{bc},\; R_{a\bar b c\bar d}$ are,
respectively, the connection and curvature of the K\"ahler
structure. The corresponding Poisson brackets are defined by the
following non-zero relations (and their complex-conjugates):
\beq
\begin{array}{c}
\{\pi_a, z^b\}=\delta^b_a,\quad
\{\pi_a,\eta^b_\alpha\}=-\Gamma^b_{ac}\eta^c_\alpha,\\
\{\pi_a,\bar\pi_b\}=i(Bg_{a\bar b}+ i R_{a\bar b c\bar
d}\eta^c_\alpha{\bar\eta}^d_\alpha), \quad \{\eta^a_\alpha,
\bar\eta^b_\beta\}= g^{a\bar b}\delta_{\alpha\beta}.
\end{array}
\eeq
The symplectic structure (\ref{ss}) becomes canonical in the
coordinates $(p_a,\chi^k)$
\begin{equation}
\begin{array}{c}
p_a=\pi_a-\frac{i}{2} \partial_a{\bf g}, \quad\chi^m_i={\rm
e}^m_b\eta^b_i:\quad \\
 \Omega_{Scan}=dp_a\wedge d z^a + d{\bar
p}_{\bar a}\wedge d{\bar z}^{\bar a} + iB g_{a\bar b}dz^a\wedge
d{\bar z}^b +d\chi^m_\alpha\wedge d{\bar\chi}^{\bar m}_\alpha,
\end{array}
\label{canonical}\end{equation} where ${\rm e}^m_a$ are the
einbeins of the K\"ahler structure: ${\rm e}^m_a\delta_{m\bar
m}{\bar{\rm e}}^{\bar m}_{\bar b} =g_{a\bar b}.$

So, in order to quantize the system, one chooses
$$ {\hat p}_a=-i\left(\frac{\partial}{\partial z^a}-
iB\frac{\partial K}{\partial z^a}\right) ,\quad
 {\hat{\bar  p}}_{\bar a}=
-i\left(\frac{\partial}{\partial {\bar z}^{\bar a}}+i
B\frac{\partial K}{\partial {\bar z}^a}\right),\quad
[{\hat\chi}^m_\alpha,{\hat{\bar\chi}}^{\bar n}_\beta]_+
=\delta^{m\bar n}\delta_{\alpha\beta}.
$$
Then, in order to  construct the system with the exact  ${\cal N}=2$
supersymmetry
 \begin{equation}
\{Q_+,Q^-\}={\cal H},
\{ Q_\pm ,Q_\pm\}=\{Q_\pm , {\cal H}\}=0,
\label{4sualg}\end{equation}
we shall search for the odd functions $Q^\pm$, which obey the equations
$\{Q^\pm,Q^\pm\}=0$ (we  restrict ourselves to the supersymmetric mechanics
whose  supercharges  are {\it linear}  in the
Grassmann variables $\eta^a_i$, $\bar\eta^{\bar a}_i$
).
In that case, the Poisson bracket $\{Q_+,Q_-\}$ yields  the ${\cal N}=2$ supersymmetric Hamiltonian.

 Let us search for the realization of supercharges
among the functions
\beq Q^\pm=\cos\lambda\;\Theta^\pm_1
+\sin\lambda\;\Theta^\pm_2\;, \eeq where \beq \Theta^+_1=\pi_a
\eta^a_1+ i\bar\partial_a W {\bar \eta}^a_2,\quad
\Theta^+_2={\bar\pi}_a\bar\eta^a_2 +i\;\partial_a W \eta^a_1,\quad
\quad \Theta^-_{1,2}={\bar\Theta}^+_{1,2}, \eeq
and $\lambda$ is
some parameter.

 Calculating the Poisson brackets of the functions,
we get
\begin{eqnarray}
&\{Q^\pm,Q^\pm\}=& i(B\sin 2\lambda\; +
2\omega\cos 2\lambda ){\cal F}_\pm ,\label{pp}\\
&  \{Q^+,Q^-\}=& {\cal H}^0_{SUSY}+ \left(B\cos 2\lambda\;
-2\omega\sin 2\lambda \right)\;{\cal F}_3/2. \label{pm}\end{eqnarray}
 Here and in
the following, we use the notation \beq {\cal H}^0_{SUSY}={\cal H}
-R_{a\bar b c\bar d}\eta^a_1\bar\eta^b_1\eta^c_2\bar\eta^d_2
-iW_{a;b}\eta^a_1\eta^b_2+ iW_{\bar a;\bar
b}\bar\eta^a_1\bar\eta^b_2+
 B \frac{{  ig_{a\bar b}\eta^a_\alpha{\bar\eta}^b_\alpha  }}{2}, \label{hosup}
\eeq
where  ${\cal H}$ denotes the oscillator Hamiltonian
(see the expression in (\ref{0})) and
\beq
{\cal F}_i= \frac{i}{2}g_{a\bar
b}\eta^a_\alpha\bar\eta^b_\beta\sigma_{(i)\alpha\bar\beta},
\quad{\cal
F}_\pm={\cal F}_1\pm{\cal F}_2.
 \eeq
%  which obey the commutation relations
%\begin{eqnarray}
%&\{{\cal F}_\pm, {\cal F}_3\}= \mp 2i{\cal F}_\pm,\quad
%\{{\cal F}_+, {\cal F}_-\}=i{\cal F}_3&\label{ff}\\
%&\{{\Theta}^\pm_\alpha, {\cal F}_\pm\}=0,\quad
%\{{\Theta}^\pm_\alpha, {\cal F}_\mp\}= \pm
%i\epsilon_{\alpha\beta}{\Theta}^\mp_\beta,\quad
%\{{\Theta}^\pm_\alpha, {\cal F}_3\}=
%\pm i{\Theta}^\pm_\alpha,&\label{thf}
%\end{eqnarray}
One has, then
\beq \{Q^\pm,
Q^\pm\}=0\Leftrightarrow B\sin\;2\lambda+2\omega\cos\;2\lambda=0,
\eeq so that
\beq \lambda=\lambda_0+(\alpha -1)\pi/2,\quad \alpha=1,2. \eeq
 Here the parameter   $\lambda_0$ is defined by the expressions\beq \cos
2\lambda_0=\frac{B/2}{\sqrt{\omega^2+(B/2)^2}},\quad \sin
2\lambda_0=-\frac{\omega}{\sqrt{\omega^2+(B/2)^2}}. \eeq
Hence, we get the following supercharges: \beq
 Q^\pm_\alpha=\cos\lambda_0\Theta^\pm_1+
(-1)^\alpha\sin\lambda_0\Theta^\pm_2, \eeq and  the pair of
 ${\cal N}=2$ supersymmetric Hamiltonians
\beq {\cal
H}^\alpha_{SUSY}= \{Q^+_\alpha,Q^-_\alpha\}=   {\cal H}^0_{SUSY}
-(-1)^\alpha
{\sqrt{\omega^2+(B/2)^2}} {\cal F}_3 \label{n4}\eeq
Notice that the supersymmetry invariance is preserved in the presence of the
constant magnetic field.

Calculating the commutators of $Q^\pm_1$ and  $Q^\pm_2$ we get
\beq \{Q^\pm_1,Q^\pm_2\}=2{\sqrt{\omega^2+(B/2)^2}} {\cal F}_\pm , \quad
\{Q^+_1,Q^-_2\}=0,\\
\eeq where the Poisson brackets between ${\cal F}_\pm $, and
$Q^\pm_\alpha$ look as follows: \beq
\begin{array}{c}
\{Q^\pm_\alpha, {\cal F}_\pm\}=0,\quad \{Q^\pm_\alpha, {\cal
F}_\mp\}= \pm\epsilon_{\alpha\beta}Q^\pm_\beta,\quad
\{Q^\pm_\alpha, {\cal F}_3\}=\pm iQ^\pm_\alpha.
\end{array}
\eeq
The whole superalgebra reads
\beq
\begin{array}{c}
\{Q^\pm_\alpha,Q^\pm_\beta\}=2\Lambda\epsilon_{\alpha\beta} {\cal
F}_\pm,\quad \{Q^\pm_\alpha,Q^\mp_\beta\}=
\delta_{\alpha\beta}{\cal H}^0_{SUSY}
-\Lambda \sigma^3_{\alpha\beta}{\cal F}_3,\\
\{Q^\pm_\alpha, {\cal F}_\pm\}=0,\quad \{Q^\pm_\alpha, {\cal
F}_\mp\}= \pm\epsilon_{\alpha\beta}Q^\pm_\beta,\quad
\{Q^\pm_\alpha, {\cal F}_3\}=\pm iQ^\pm_\alpha ,\\
\{{\cal F}_\pm,{\cal F}_\mp\}=i{\cal F}_3, \quad\{{\cal
F}_\pm,{\cal F}_3\}=\pm i{\cal F}_\pm \;.
\end{array}
\eeq
where
\beq
\Lambda={\sqrt{\omega^2+(B/2)^2}}.
\eeq
Let us notice  the $\omega$ and $B$ appear in this superalgebra
in a symmetric way,
via the factor ${\sqrt{\omega^2+(B/2)^2}}$.

This superalgebra could be represented in a bit more convenient form, if introduce
\beq
S^\pm_1\equiv Q^\pm_1,\quad S^\pm_2=Q^\mp_2.
\eeq
In this notation, it reads
\beq
\{S^\pm_\alpha, S^\mp_\beta\}=\delta_{\alpha\beta}{\cal H}+
\Lambda\sigma^i_{\alpha\beta}{\cal F}_i,\quad
\{S^\pm_\alpha, {\cal F}_i\}=\pm\imath\sigma^i_{\alpha\beta}S^\pm_{\beta}, \quad
\{{\cal F}_i, {\cal F}_{j}\}=\epsilon_{ijk}{\cal F}_k.
\eeq
All other commutators vanish.\\

This is precisely the weak supersymmetry algebra  considered by A. Smilga \cite{smilga}.
In the particular case $\omega=0$
it yields the ${\cal N}=4$ supersymmetric mechanics broken
by a constant magnetic field.
\\

{\bf Remark 1.} In the case of the oscillator on $\DC^n$ we can smoothly
relate the above supersymmetric oscillator  with a ${\it N=4}$ oscillator,
if choose
\beq
K=\cos{\gamma}\; z\bar z+\sin{\gamma}\;( z^2+ \bar z^2)/2
\; ,\qquad \gamma \in [0,\pi/2].
\eeq
Hence,
\beq
{\cal H}=\pi\bar\pi+\omega_0^2 z\bar z
+\sin2\gamma\; \omega^2_0 (z^2+\bar z^2)/2\;,
\eeq
i.e. for $\gamma=0,\pi /2$ we have a standard harmonic oscillator, while
for $\gamma\neq 0, \pi/2$ we get the anisotropic one, which is
 equivalent to two sets of $n$ one-dimensional oscillators
with frequencies  $\omega_0{\sqrt{1\pm\sin2\gamma}}$.
The frequency $\omega$ appearing in the superalgebra,
is of the form: $\omega=\omega_0\cos\gamma $.\\

{\bf Remark 2.} In a similar way, we can consider the supersymmetrization
of the two-dimensional noncommutative oscillator
in the constant magnetic field \cite{ncqm}.
For this purpose, we define the Poisson brackets  \cite{pouliot}
\beq
\{\pi,z\}=1,\quad \{z,\bar z\}=i\theta,\quad \{\pi,\bar\pi\}=iB,\quad
\{\eta_\alpha,\bar\eta_\beta\}=\delta_{\alpha\beta}
\eeq
and choose the following supercharges:
\beq
S^+_\alpha=\pi\eta_\alpha+ i \epsilon_{\alpha\beta}\bar z\bar\eta_{\beta},
\quad  S^-_\alpha=\bar S^+_\alpha.
\eeq
Calculating the Poisson brackets, we get the same superalgebra as above,
with the following Hamiltonian and $\Lambda$ parameter:
\beq
{\cal H}^0_{SUSY}=\pi\bar\pi+\omega^2 z\bar z-i\omega\eta_1\eta_2 +i\omega\bar\eta_1\bar\eta_2
+\frac{i}{2}
{(B+\theta\omega^2)}\eta_\alpha\bar\eta_\alpha,\qquad
\Lambda=B-\theta\omega^2.
\eeq
Hence, the exact  ${\cal N}=4$ supersymmetry is realized
only under the choice of the parameter
\beq
{\Lambda}=0\quad\Leftrightarrow \quad B=\theta\omega^2.
\eeq
This is not surprising, since under this choice of $\Lambda$ the underlying system
 is equivalent to the isotropic oscillator \cite{ncqm}.\\
%\section{Discussion}

{\bf Acknowledgments}
We are indebted to Armen Yeranyan for useful
discussions.
The work of S.B. was supported in part
by the European Community's Human Potential
Programme under
contract HPRN-CT-2000-00131 Quantum Spacetime,
the INTAS-00-00254 grant and the
NATO Collaborative Linkage Grant PST.CLG.979389.
The work of A.N. was supported by grants INTAS 00-00262  and ANSEF  PS81.

\end{document}